\documentclass[11pt]{article}
\usepackage{amssymb}
\usepackage{graphics}
\usepackage{epsfig}
\usepackage{a4wide}
\usepackage[normal]{caption2}
\newcommand{\sss}{\scriptscriptstyle}

\begin{document}

\title{ NLO Supersymmetric QCD Corrections to $t \bar t h^0$ Associated Production at Hadron Colliders  }
\vspace{3mm}

\author{{ Wu Peng$^{2}$, Ma Wen-Gan$^{1,2}$, Hou Hong-Sheng$^{2}$, Zhang Ren-You$^{2}$,
Han Liang$^{2}$ and Jiang Yi$^{2}$}\\
{\small $^{1}$CCAST (World Laboratory), P.O.Box 8730, Beijing, 100080, People's Republic of China} \\
{\small $^{2}$Department of Modern Physics, University of Science and Technology of China (USTC),}\\
{\small       Hefei, Anhui 230027, People's Republic of China} }
\date{}
\maketitle` \vskip 12mm

\begin{abstract}
We calculate NLO QCD corrections to the lightest neutral Higgs
boson production associated with top quark pair at hadron
colliders in the minimal supersymmetric standard model(MSSM). Our
calculation shows that the total QCD correction significantly
reduces its dependence on the renormalization/factorization scale.
The relative correction from the SUSY QCD part approaches to be a
constant, if either $M_S$ or $m_{\tilde{g}}$ is heavy enough. The
corrections are generally moderate(in the range of few percent to
$20\%$) and under control in most of the SUSY parameter space. The
relative correction is obviously related to $m_{\tilde{g}}$, $A_t$
and $\mu$, but not very sensitive to $\tan\beta$, $M_S$ at both
the Tevatron and the LHC with our specified parameters.

\end{abstract}

\vskip 5cm


\vfill \eject

\renewcommand{\theequation}{\arabic{section}.\arabic{equation}}
\renewcommand{\thesection}{\Roman{section}.}
\newcommand{\nb}{\nonumber}
\renewcommand{\captionlabeldelim}{.}
\renewcommand{\captionlabelfont}{\bfseries}
\renewcommand{\figurename}{Fig.}
\setlength{\abovecaptionskip}{-10pt}

\makeatletter      
\@addtoreset{equation}{section}
\makeatother       

\par
\section{Introduction}
\par
One of the major objectives of future high-energy experiments is
to search for scalar Higgs particles and investigate the symmetry
breaking mechanism of the electroweak interactions. In the
standard model (SM) \cite{sm}, one doublet of complex scalar
fields is introduced to spontaneously break the symmetry, leading
to a single neutral Higgs boson $h^0$. But there exists the
problem of the quadratically divergent contributions to the
corrections to the Higgs boson mass. This is the so-called
naturalness problem. One of the hopeful methods, which can solve
this problem, is the supersymmetric (SUSY) extension to the SM. In
these extension models, the quadratic divergences of the Higgs
boson mass can be cancelled by loop diagrams involving the
supersymmetric partners of the SM particles exactly. The most
attractive and simplest supersymmetric extension of the SM is the
minimal supersymmetric standard model (MSSM)\cite{mssm-1,mssm-2}.
In this model, there are two Higgs doublets $H_1$ and $H_2$ to
give masses to up- and down-type fermions. The Higgs sector
consists of three neutral Higgs bosons, one $CP$-odd particle
($A^0$), two $CP$-even particles ($h^0$ and $H^0$), and a pair of
charged Higgs bosons ($H^{\pm}$).
\par
However, these Higgs bosons haven't been directly explored
experimentally until now. The published experimental lower mass
bounds for the Higgs bosons presented by LEP experiments are:
$M_{h^0}\!>\!114.4$~GeV (at $95\%$ CL) for the SM Higgs boson, and
for the MSSM bosons $M_{h^0}\!>\!91.0$~GeV and
$M_{A^0}\!>\!91.9$~GeV (at $95\%$ CL, $0.5\!<\!\tan\beta\!<\!2.4$
excluded). The SM fits to precision electroweak data
\cite{lepewwg} indirectly set a limitation of the light Higgs
boson, $M_{h^0}\!<\!200$~GeV, while there should has a scalar
Higgs boson lighter than about 130~GeV in MSSM. \cite{Heinemeyer}.
This lightest Higgs boson with mediate mass is certainly in the
exploring mass range of the present and future colliders, such as
the Tevatron Run II, LHC and LC. At a LC the cross section for
$e^{+}e^{-} \rightarrow t \bar{t} h$ is small, about $1~{\rm fb}$
for $\sqrt{s} = 500~{\rm GeV}$ and $m_h=100~ {\rm GeV}$ \cite{ee,
ee2}. But it has a distinctive experimental signature and can
potentially be used to measure the top quark Yukawa coupling in
the intermediate Higgs mass region at a LC with very high
luminosity. S. Dawson and L. Reina calculated the NLO QCD
corrections to $e^+e^- \to t\bar t h^0$ process at LC's in Ref.
\cite{ee1}. And in references \cite{tthc1,tthc2,tthc3} the SM
electroweak corrections to the process $e^+e^- \to t\bar t h^0$
are calculated. H. Chen et al., have studied the QCD and
electroweak corrections to the process $\gamma\gamma \to t\bar t
h^0$ in the SM at LC's\cite{HChen}. All these works show that the
evaluation of radiative corrections is a crucial task for all
accurate measurements of $t\bar t h^0$ production process.
\par
There are various channels which can be exploited to search for
the Higgs boson $h^0$ with intermediate mass at TeV energy scale
hadron colliders, such as gluon-gluon fusion Higgs boson
production($gg\rightarrow h^0$), the associated production with a
weak intermediate boson ($q q' \to W h^0, Z h^0$). Recently, the
production channels $pp/p\bar p \to t\bar{t} h^0+X$ attracted the
physicist's attentions, because these channels offer a spectacular
signature ($W^+W^- b \bar b b \bar b$)\cite{Goldstein} and
provides a possibility in probing the Yukawa
coupling\cite{ppqcd1,ppqcd2}. The total cross section for
$pp/p\bar p \to t\bar{t} h^0+X$ at tree level and NLO QCD
corrections in the SM have been studied in
Refs.\cite{ppqcd1,ppqcd2, Kunszt, ppbar3, eq1}.
\par
The supersymmetric (SUSY) electroweak corrections to the $e^+e^-
\to t\bar t h^0$ process can be over ten percent for favorable
parameter values\cite{CSLi}. In Ref.\cite{zhu}, it was found that
the SUSY QCD interactions by exchanging gluinos and squarks can
impact on the Yukawa coupling vertex in the process $e^+ e^-\to t
\bar{t} h^0$ at LC. At $pp/p\bar p$ hadron colliders with a
center-of-mass energy of TeV scale, the dominated contributions to
$t\bar t h^0$ production are from subprocesses $q\bar{q},gg \to
t\bar{t} h^0$. To these high energy $t\bar{t} h^0$ production
processes, the SUSY radiative corrections, especially the SUSY QCD
corrections, may be remarkable.
\par
In this paper, we calculated the cross section for the associated
production of the Higgs boson with top quark pair in the MSSM at
hadron colliders including the NLO QCD corrections. In section 2,
we present the calculations of the leading order cross sections to
$pp/p\bar p \to t\bar t h^0+X$ in the MSSM. In section 3, we
present the calculations of the ${\cal O}(\alpha_{{\rm s}}^3)$ QCD
corrections to $pp/p\bar p \to t\bar t h^0+X$ in the MSSM. The
numerical results and discussions are presented in section 4.
Finally, a short summary is given.

\section{The leading order cross sections}
\par
The Feynman diagrams at leading order(LO) for the subprocess
\begin{equation}
q(p_1)\bar q(p_2) \to t(k_1)\bar t(k_2) h^0(k_3),
\end{equation}
in the MSSM are plotted in Fig.1. They are s-channel, gluon
exchange diagrams with Higgs boson radiation off top-quark and
anti-top-quark, respectively. The process
\begin{equation}
g(p_1)g(p_2) \to t(k_1)\bar t(k_2) h^0(k_3),
\end{equation}
at the tree level in the MSSM are described by the Feynman
diagrams of Fig.2. The LO Feynman diagrams for both subprocesses
in the MSSM are the same with their corresponding ones in the SM.
In above two channels we use $p_{1,2}$ and $k_{1,2,3}$ to
represent the four-momenta of the incoming partons and the
outgoing particles, respectively. Because of the small mass of u-
and d-quark, we neglect the diagrams which involve $h^0-u-\bar{u}$
and $h^0-d-\bar{d}$ Yukawa vertexes.
\begin{figure}[htp]
\includegraphics*[0,570][380,630]{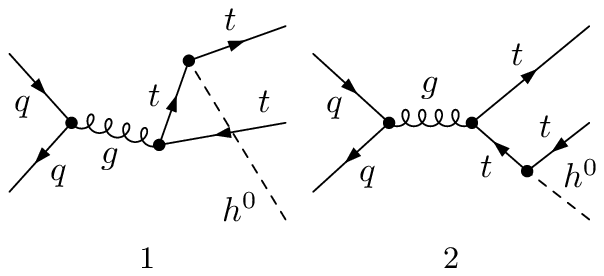}
\center{\caption{The tree-level Feynman diagrams for the $q\bar{q}
\to t\bar{t}h$ subprocess.}}
\end{figure}
\begin{figure}[htp]
\includegraphics*[70,460][500,630]{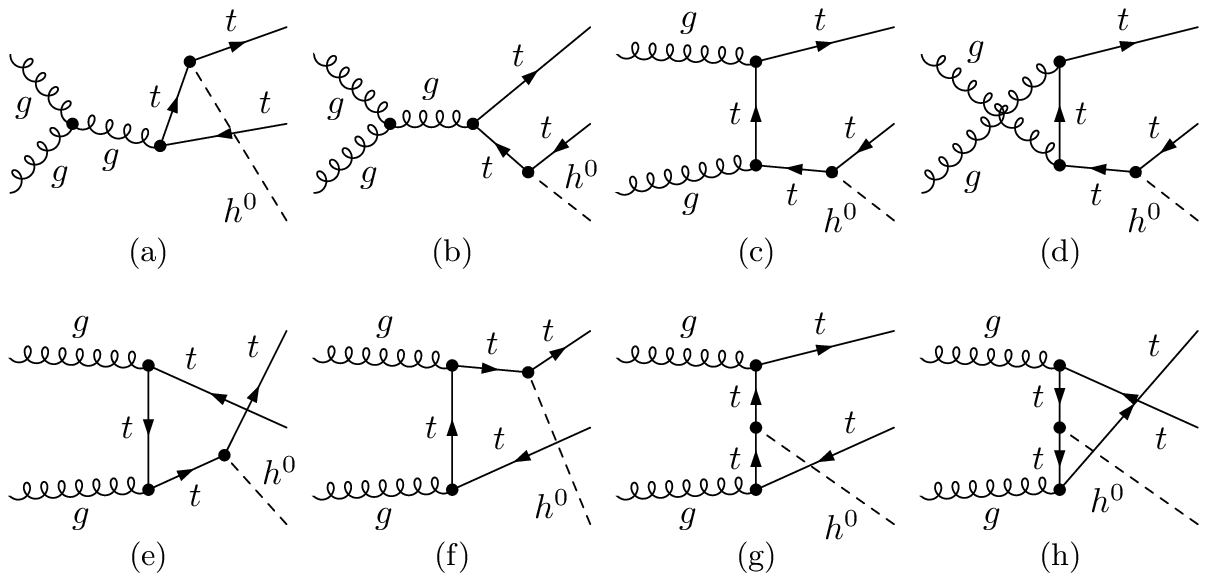}
\center{\caption{The tree-level Feynman diagrams for the $gg \to
t\bar{t}h$ subprocess.}}
\end{figure}
\par
The explicit expression for the amplitudes of subprocess $q\bar q
\to t\bar t h^0$ at tree level can be written as:
\begin{equation}
M_{LO}^{q\bar{q}}= A
   \bar{v}_k(p_2)\gamma_{\mu}u_i(p_1)\frac{ig_{\mu\nu}}{\hat
s}\bar{u}_j(k_1)\frac{\rlap/k_1+\rlap/k_3+m_t}
   {(k_1+k_3)^2-m_t^2}\gamma_{\nu}v_l(k_2)T^a_{ik}T^b_{jl}+(k_1 \leftrightarrow
k_2)
\end{equation}
where $\hat{s}=(p_1+p_2)^2$, $T^a$ is the $SU(3)$ color matrix,
$A=g^2_sY_{tth}^{(SUSY)}$. $g_s$ is the strong coupling constant.
$Y_{tth}^{(SUSY)}$ is the Yukawa coupling between Higgs boson and
top quarks in the MSSM. As we know the $h^0-t-\bar t$ Yukawa
coupling in the SM $Y_{tth}^{(SM)}$ is expressed by
\begin{eqnarray}
Y_{tth}^{(SM)}=-i g_w \frac{m_t}{2 m_W},
\end{eqnarray}
But in the MSSM, $Y_{tth}^{(SUSY)}$ is given as
\begin{eqnarray}
Y_{tth}^{(SUSY)}=-i g_w \frac{m_t}{2 m_W} \frac{{\rm cos}
\alpha}{{\rm sin} \beta},
\end{eqnarray}
where $\alpha$ is the mixing angle which leads to the physical
Higgs boson eigenstates $h^0$ and $H^0$. The angle $\beta$ is
defined as $\tan\beta=v_2/v_1$, where $v_1$ and $v_2$ are the
vacuum expectation values.

\par
According to the different topologies of Feynman diagrams, the
explicit expression for the amplitudes of subprocess $gg\to t\bar
t h^0$ in the MSSM at tree level can be divided into three parts.
\begin{equation}
M_{tree}^{gg}= M_{tree}^{gg1}+M_{tree}^{gg2}+M_{tree}^{gg3}
\end{equation}
where $M_{tree}^{gg1}$, $M_{tree}^{gg2}$ and $M_{tree}^{gg3}$
correspond to the amplitudes of Fig.2(a-b), Fig.2(c-f) and
Fig.2(g-h), respectively. For the amplitude parts
$M_{tree}^{ggi}(i=1,2,3)$, we have the expressions as:
\begin{equation}
M_{tree}^{gg1}=
Af^{abc}T^c_{ij}\bar{u}_j(k_1)\frac{\rlap/\epsilon^{\sss\mu}_1\rlap/\epsilon^{\sss\nu}_2}{\hat
s}[2p^{\nu}_1g^{\lambda\mu}+(p_2-p_1)^{\lambda}g^{\mu\nu}-2p^{\mu}_2g^{\nu\lambda}]\frac{-\rlap/k_2-\rlap/k_3+m_t}
   {2k_2 \cdot k_3+m_{h^{\sss 0}}^2}v_i(k_2)+(k_1 \leftrightarrow
k_2)
\end{equation}
\begin{equation}
M_{tree}^{gg2}=
-iAT^a_{ik}T^b_{kj}\bar{u}_j(k_1)\frac{\rlap/k_3+\rlap/k_1+m_t}{2k_3
\cdot
k_1+m_{h^0}^2}\rlap/\epsilon_2\frac{-\rlap/k_2+\rlap/p_1+m_t}{-2k_2
\cdot p_1}\rlap/\epsilon_1v_i(k_2)+\left(%
\begin{array}{c}
  p_1 \leftrightarrow p_2, \epsilon_1
\leftrightarrow \epsilon_2 \\
  p_1 \leftrightarrow p_2, \epsilon_1
\leftrightarrow \epsilon_2,k_1 \leftrightarrow k_2 \\
  k_1 \leftrightarrow k_2 \\
\end{array}%
\right)
\end{equation}
\begin{equation}
M_{tree}^{gg3}=
-iAT^a_{ik}T^b_{kj}\bar{u}_j(k_1)\rlap/\epsilon_2\frac{\rlap/k_1-\rlap/p_2+m_t}{-k_1
\cdot p_2}\frac{-\rlap/k_2+\rlap/p_1+m_t}{-k_2 \cdot
p_1}\rlap/\epsilon_1v_i(k_2)+(p_1 \leftrightarrow p_2, \epsilon_1
\leftrightarrow \epsilon_2)
\end{equation}
where $\epsilon^{\mu}_1$ and $\epsilon^{\nu}_2$ are the
polarization four-vectors of the incoming gluons. The SU(3)
structure constants are given by $f_{abc}$. Then the lowest order
cross sections for the subprocesses $q\bar q,gg \to t\bar t h^0$
in the MSSM are obtained by using the following formula:
\begin{eqnarray}
\hat\sigma_{{LO}}^{q\bar{q},gg} =
\frac{1}{2|\vec{k}_1|\sqrt{\hat{s}}}\int {\rm d}
\Phi_3\overline{\sum} |M_{tree}^{q\bar q,gg}|^2
\end{eqnarray}
where ${\rm d} \Phi_3$ is the three-body phase space element. The
summation is taken over the spins and colors of initial and final
states, and the bar over the summation recalls averaging over the
spins and colors of initial partons. The LO total cross section of
$pp/p\bar{p} \to t\bar{t} h^0+X$ can be expressed as:
\begin{equation}
\label{eq:sigma_nlo} \sigma_{LO}(pp/p\bar p\to t\bar t h^0+X)
=\sum_{ij}\int dx_1 dx_2 {G}_i^p(x_1,\mu) {G}_j^{p/\bar
p}(x_2,\mu) {\hat \sigma}^{ij}_{LO}(x_1,x_2,\mu)\,\,\,,
\end{equation}
where ${\hat \sigma}^{ij}_{LO}(ij=q\bar q,gg)$ is the LO
parton-level total cross section for incoming $i$ and $j$ partons,
$G_i^{p/\bar p}$'s are the LO parton distribution functions (PDF)
with parton $i$ in a proton/antiproton.
\par
From the above deduction, we can see that the ratio between the
tree level cross sections of subprocess $q\bar q(gg) \to t\bar t
h^0$ in the SUSY model and the SM, is written as
\begin{equation}
\frac{\hat\sigma_{{\sss LO}}^{(SUSY)}(q\bar q,gg \to t\bar t
h^0)}{\hat\sigma_{{\sss LO}}^{(SM)}(q\bar q,gg \to t\bar t
h^0)}=\frac{{\rm cos}^2 \alpha}{{\rm sin}^2 \beta}.
\end{equation}

\vskip 5mm
\section{NLO QCD Corrections in the MSSM}

In the calculation of the NLO QCD corrections in the MSSM, we
adopt the dimensional regularization in $D=4-2\epsilon$ dimensions
to isolate the ultraviolet(UV), infrared(IR) and collinear
singularities. Renormalization and factorization are performed in
the modified minimal substraction($\overline{MS}$) scheme, and the
wave functions of the external fields, and top quark's mass in
propagators and in the Yukawa couplings are renormalized in the
on-shell(OS) scheme. We divide the ${\cal O}(\alpha_s^3)$ QCD
correction to the subprocess $q\bar q(gg) \to t\bar th^0$ in the
MSSM into two parts. One is the so-called SM-like QCD correction
part, another is SUSY-QCD correction part arising from virtual
gluino/squark exchange contributions. Then the total NLO QCD
corrections and relative corrections in the MSSM can be expressed
as
\begin{eqnarray}
\Delta \hat{\sigma}_{NLO}^{(q\bar q,gg)}=\Delta
\hat{\sigma}_{SM-like}^{(q\bar q,gg)}+\Delta
\hat{\sigma}_{SQCD}^{(q\bar q,gg)},~~~~\hat{\delta}^{(q\bar
q,gg)}=\hat{\delta}_{SM-like}^{(q\bar q,gg)}+\hat{\delta}
_{SQCD}^{(q\bar q,gg)}.
\end{eqnarray}
where we define the relative correction as
$\hat{\delta}=\frac{\Delta
\hat{\sigma}_{NLO}}{\hat{\sigma}_{LO}}$. The NLO SM-like QCD
correction part(relative correction part) in the MSSM has
following relation with the NLO SM QCD one
\begin{eqnarray}
\Delta \hat{\sigma}_{SM-like}=(\frac{\cos\alpha}{\sin
\beta})^2\Delta \hat{\sigma}_{SM},
~~~\hat{\delta}_{SM-like}=\hat{\delta}_{SM}.
\end{eqnarray}
In our calculation we introduce the following counterterms.
\begin{eqnarray}
\label{defination of renormalization constants}
m_t & \to & m_t+\delta m_t,~~~~g_s \to g_s+\delta g_s    \nb \\
t_L  & \to & \left( 1+\frac{1}{2}\delta Z_{L}^t\right)t_L,~~~~
t_R \ \to \ \left( 1+\frac{1}{2}\delta Z_{R}^t\right)t_R \nb \\
u_L  & \to & \left( 1+\frac{1}{2}\delta Z_{L}^u\right)u_L,~~~~
u_R \ \to \ \left( 1+\frac{1}{2}\delta Z_{R}^u\right)u_R \nb \\
d_L  & \to & \left( 1+\frac{1}{2}\delta Z_{L}^d\right)d_L,~~~~
d_R \ \to \ \left( 1+\frac{1}{2}\delta Z_{R}^d\right)d_R \nb \\
G_{\mu} & \to & (1+ \frac{1}{2}\delta Z_g)G_{\mu},
\end{eqnarray}
where $g_s$ denotes the strong coupling constant, $t,~u,~d$ and
$G_{\mu}$ denote the fields of top-, up-, down-quark and gluon.
The definitions and the explicit expressions of these
renormalization constants can be found in Ref. \cite{COMS}. For
the renomalization of the QCD coupling constant $g_{\rm s}$, we
use the $\overline{MS}$ scheme except that the divergences
associated with the colored SUSY particle loops are subtracted at
zero momentum\cite{gs}. Since we have $\delta g=\delta
g^{(SM-like)}+\delta g^{(SQCD)}$, the terms should be obtained as
\begin{eqnarray}
&& \frac{\delta g^{(SM-like)}_s}{g_s}=
-\frac{\alpha_s(\mu_r^2)}{4\pi}
\left[\frac{\beta^{(SM-like)}_0}{2}\frac{1}{\bar{\epsilon}}
+\frac{1}{3}\ln\frac{m_{t}^2} {\mu_{r}^2}\right],
\end{eqnarray}
\begin{eqnarray}
&& \frac{\delta g^{(SQCD)}_s}{g_s}=
-\frac{\alpha_s(\mu_r^2)}{4\pi}
\left[\frac{\beta^{(SQCD)}_0}{2}\frac{1}{\bar{\epsilon}}
+\frac{N}{3}\ln\frac{m_{\tilde{g}}^2} {\mu_{r}^2}
+\sum_{U=u,c,t}^{i=1,2}\frac{1}{12}\ln\frac{m_{\tilde{U_i}}^2}{\mu_{r}^2}
+\sum_{D=d,s,b}^{j=1,2}\frac{1}{12}\ln\frac{m_{\tilde{D_j}}^2}{\mu_{r}^2}\right],
\end{eqnarray}
where
\begin{eqnarray}
&& \beta^{(SM-like)}_0=\frac{11}{3}N-\frac{2}{3}n_f-\frac{2}{3} ,
~~~~\beta^{(SQCD)}_0=-\frac{2}{3}N-\frac{1}{3}(n_f+1),
\end{eqnarray}
$N=3$ ,$n_f=5$ and $1/\bar{\epsilon}=1/\epsilon_{UV} -\gamma_E
+\ln(4\pi)$. The summation is taken over the indexes of squark and
generation. The $\overline{MS}$ scheme violates supersymmetry
explicitly, and the $q\tilde{q}\tilde{g}$ Yukawa coupling
$\hat{g}_s$, which should be the same with the $qqg$ gauge
coupling $g_s$ in the supersymmetry, takes a finite shift at
one-loop order as shown in Eq.(\ref{shift}) \cite{shiftgs}.
\begin{eqnarray}
\label{shift} && \hat{g}_s = g_s [1
+\frac{\alpha_s}{8\pi}(\frac{4}{3}N - C_F)],
\end{eqnarray}
with $N=3$ and $C_F=4/3$. In our numerical calculation we take
this shift between $\hat{g}_s$ and $g_s$ into account.

\par
Actually, the calculation of the NLO SM-like QCD corrections in
the MSSM for the subprocesses $q\bar q,gg \to t\bar{t}h^0$ is the
same as that of the NLO SM QCD corrections in
Refs.\cite{ppqcd1,ppqcd2}, except their numerical results
satisfied the relations shown in Eq.(3.2).
\par
The NLO SUSY-QCD contribution part to the $q\bar q(gg) \to
t\bar{t}h^0$ subprocess comes from the one-loop diagrams involving
virtual gluino/squark exchange. For demonstration, we show the
pentagon diagrams which contribute to the NLO SUSY-QCD correction
part for the subprocesses $q\bar q \to t\bar th^0$ and $gg \to
t\bar th^0$ in Fig.3, where the upper indexes $s,t,u=1,2$. Because
there is no massless particle in the loop, all these diagrams with
gluino/squark loop are IR finite. The pentagon and box diagrams in
SUSY-QCD part are UV finite, but the self-energy and vertex
diagrams in this part contain UV divergences. That is renormalized
by the proper related counterterms defined in Eq.(3.3).
\begin{figure}[htp]
\includegraphics*[80,270][580,630]{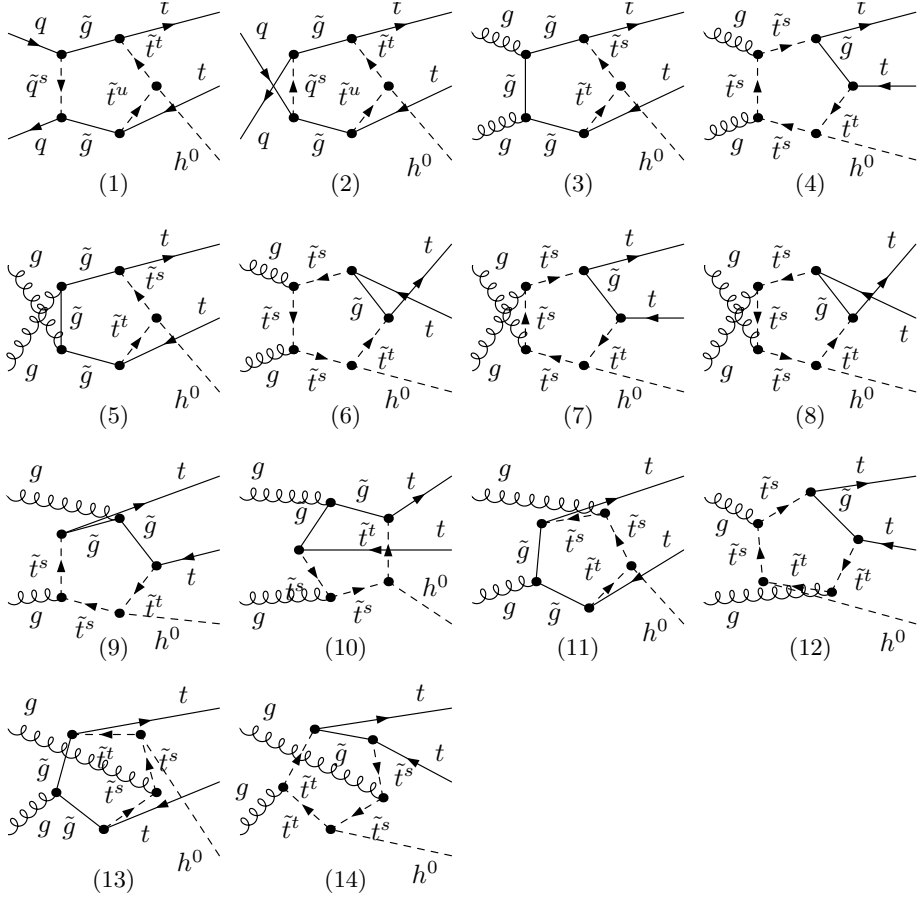}
\center{\caption{The pentagon diagrams for the $q\bar{q} \to
t\bar{t}h^0$ and $gg\to t\bar{t}h^0$ subprocess.}}
\end{figure}
\par
The ${\cal O}(\alpha_{{\rm s}}^3)$ supersymmetric QCD correction
part of the cross section in the MSSM to the subprocesses $q\bar
q,gg \to t\bar t h^0$ can be expressed as
\begin{eqnarray}
\Delta\hat\sigma_{SQCD}^{(q\bar q,gg)} =
\frac{1}{2|\vec{k}_1|\sqrt{\hat s}}\int {\rm d}
\Phi_3\overline{\sum} 2{\rm Re}\left( {\cal M}_{tree}^{(q\bar
q,gg)} {\cal M}_{{\rm SQCD}}^{(q\bar q,gg)\dag} \right),
\end{eqnarray}
where ${\cal M}_{tree}^{(q\bar q,gg)}$ are the Born amplitudes for
$q\bar{q},gg \to t\bar{t}h^0$ subprocesses, and ${\cal M}_{{\rm
SQCD}}^{(q\bar q,gg)}$ are the renormalized amplitudes of all the
one-loop Feynman diagrams involving virtual gluino/squark.
\par
In the calculations of loop diagrams we adopt the definitions of
one-loop integral functions of Ref.\cite{s14}. The Feynman
diagrams and the relevant amplitudes are generated by {\it
FeynArts} 3\cite{FA3}, and the Feynman amplitudes are subsequently
reduced by {\it FormCalc32}. The phase space integration is
implemented by using Monte Carlo technique. The numerical
calculations of integral functions are implemented by using
developed {\it LoopTools}.
\par
We write the NLO QCD corrected parton-level total cross section
${\hat \sigma}_{NLO}^{ij}( x_1,x_2,\mu)$ as:
\begin{eqnarray}
\label{eq:sigmahat_nlo} {\hat \sigma}_{NLO}^{(q\bar q,gg)}
&\equiv&{\hat \sigma}_{LO}^{(q\bar q,gg)}+ \Delta{\hat
\sigma}_{NLO}^{(q\bar q,gg)}\,\,\,,\nonumber
\end{eqnarray}
\par
The NLO QCD corrected total cross section of $pp/p\bar{p} \to
t\bar{t} h^0+X$ in the MSSM can be expressed as:
\begin{equation}
\label{eq:sigma_nlo} \sigma_{NLO}(pp/p\bar p\to t\bar t h^0)
=\sum_{ij}\int dx_1 dx_2 {G}_i^p(x_1,\mu) {G}_j^{p/\bar
p}(x_2,\mu) {\hat \sigma}^{(ij)}_{NLO}(x_1,x_2,\mu)\,\,\,,
\end{equation}
where ${\hat \sigma}^{(ij)}_{\sss NLO}(ij=q\bar q,gg)$ is the NLO
QCD corrected parton-level total cross section for incoming $i$
and $j$ partons, and $ G_i^{p/\bar p}$ are the NLO parton
distribution functions (PDF) for parton $i$ in a
proton/antiproton. The equation include two channels:
$q\bar{q},gg\to t\bar{t}h^0$. In our calculations, we choose the
factorization scale equals the renormalization scale, i.e.,
$\mu_f=\mu_r=Q$. The partonic center-of-mass energy squared, $\hat
s$, is given in terms of the total hadronic center-of-mass energy
squared $\hat s=x_1 x_2 s$.

\section{ Numerical Results and Discussion}

In our numerical calculation, we adopt the MRST NLO parton
distribution function\cite{MRST} and the 2-loop evolution of
$\alpha_s(\mu^2)$ to evaluate the hadronic NLO QCD corrected cross
sections, while for the hadronic LO cross sections we use the MRST
LO parton distribution function and the one-loop evolution of
$\alpha_s(\mu^2)$. We take the SM parameters as $\alpha_{{\rm
ew}}(m_Z^2)^{-1} = 127.918$, $m_W = 80.423~GeV$, $m_Z =
91.188~GeV$, $m_t = 174.3~GeV$, $m_u = m_d =
66~MeV$\cite{databook}. There we use the effective values of the
light quark masses ($m_u$ and $m_d$) which can reproduce the
hadron contribution to the shift in the fine structure constant
$\alpha_{\rm ew}(m_Z^2)$\cite{leger}. The other relevant
parameters, such as mixing angle of the Higgs fields $\alpha$ and
masses of the lightest Higgs boson, gluino, stop-quarks, are
obtained by adopting the FormCalc package, except otherwise
stated. The input parameters for the FromCalc program are $M_S$,
$M_2$, $A_t$, $m_{A^0}$, $\mu$ and $\tan\beta$. The related
parameters for the MSSM Higgs sector are obtained from the CP-odd
mass $m_{A^0}$ and $\tan\beta$ with the constraint $\tan\beta \ge
2.5$. In the program the grand unification theory(GUT) relation
$M_1 = (5/3)\tan^2 \theta_W M_2$ is adapted for simplification and
the gluino mass $m_{\tilde g}$ is evaluated by $m_{\tilde
g}={\alpha_s(Q)}/{\alpha_{ew}(m_Z)}\sin^2 \theta_W M_2$. For the
sfermion sector, the relevant input parameters are $M_S$, $A_f$
and $\mu$, and there we take the assumptions of
$M_Q=M_U=M_D=M_E=M_L=M_S$ and the soft trilinear couplings for
sfermions $\tilde{q}$ and $\tilde{l}$ being equal, i.e.,
$A_q=A_l=A_f$.
\par
We present the dependence of the cross section on the
renormalization/factorization scale $Q/Q_0$ in Fig.4(a-b) for the
Tevatron and the LHC separately, where we denote
$Q_0=m_t+m_{h^0}/2$ and the input parameters are taken as
$A_t=800~GeV$, $M_S=400~GeV$, $M_2=110~GeV$, $m_{A^0}=270~GeV$,
$\mu=-200~GeV$ and $\tan\beta=6$. With these input parameters, we
get all the other supersymmetric parameters, among them
$\cos\alpha=0.954$, $m_{h^0}=120~GeV$,
$m_{\tilde{t}_1}=207.75~GeV$ and $m_{\tilde{t}_2}=577.63~GeV$, but
the value of $m_{\tilde{g}}$ is a function of the energy scale
$Q$($m_{\tilde{g}}(Q_0)=317.9~GeV$). In order to show the cross
section dependence on the renormalization/factorization scale, we
fix $m_{\tilde{g}}=300~GeV$ in Fig.4(a-b). There we plot the
curves for cross sections $\sigma_{LO}$, $\sigma_{NLO}$ and
$\sigma_{NLO}^{SM-like}$ of the processes $p\bar p/pp \to t\bar
th^0+X$. The notations $\sigma_{NLO}$ and $\sigma_{NLO}^{SM-like}$
represent the cross sections involving complete QCD and SM-like
QCD corrections. Fig.4(a) shows that the NLO QCD contributions to
the process $p\bar p \to t\bar th^0+X$ in the MSSM at the
Tevatron, in which the dominant subprocess is $q\bar{q} \to
t\bar{t} h^0$, has a negative NLO QCD corrections near the
position of $Q=Q_0$. While Fig.4(b) shows that the NLO QCD
contributions to the process $pp \to t\bar th^0+X$ in the MSSM at
the LHC, in which the dominant subprocess is $gg \to t\bar{t}
h^0$, will give positive corrections near the position of $Q=Q_0$.
Here we should note that if $Q$ goes down to a very low value,
i.e., $Q<<Q_0$, large logarithmic corrections spoil the
convergence of perturbation theory in the proton-antiproton
colliding energy of the Tevatron. That can be seen from our
numerical results for the Tevatron. It shows that the total NLO
QCD corrected cross section $\sigma_{NLO}$ in the MSSM tends to
have a negative value when $Q \to 0$\cite{ppqcd2}. From
Fig.4(a-b), we can conclude that the dependence of the NLO QCD
corrected cross section $\sigma_{NLO}$ on the scale $Q$ is
significantly reduced comparing with $\sigma_{LO}$, and is
slightly weakened comparing with $\sigma_{NLO}^{SM-like}$.
\par
In the following calculation, we fixed the value of the
renormalization/factorization scale being $Q_0$. In Fig.5(a) and
(b) we show the LO and total NLO QCD cross sections $\sigma_{LO}$
and $\sigma_{NLO}$ in the MSSM as the functions of
$\tan\beta$$(m_{h^0})$ at the Tevatron and the LHC respectively,
taking $A_t=800~GeV$, $M_S=400~GeV$, $M_2=110~GeV$,
$m_{A^0}=270~GeV$ and $\mu=-200~GeV$. The corresponding relative
corrections $\delta$ of both cross sections versus
$\tan\beta$$(m_{h^0})$, where the relative correction is defined
as $\delta=\frac{\sigma_{NLO}-\sigma_{LO}}{\sigma_{LO}}$, are
plotted in Fig.5(c). From these figures, we can see that the cross
sections $\sigma_{NLO}$ and $\sigma_{LO}$ decrease rapidly as
$\tan\beta$ varies in the range from $2$ to $10$, and then goes
down very slowly when $\tan\beta$ changes from 10 to 40. We can
read from Fig.5(a-b) that when $\tan\beta$ increases from 2 to 40,
the total NLO QCD corrected cross section $\sigma_{NLO}$ in the
MSSM decreases roughly from $9.1~fb$ and $1078~fb$ to $5.2~fb$ and
$641~fb$ for the Tevatron and the LHC, respectively. The two
curves of relative corrections $\delta$ for the Tevatron and the
LHC in Fig.5(c) look like rather stable when $\tan\beta$ runs from
2 to 50. We can read from this figure that the NLO QCD relative
correction values in the MSSM at the Tevatron and the LHC are
generally about $-17\%$ and $26\%$ in these varying range of
$\tan\beta$, respectively.
\par
In Fig.6, we show the relative NLO QCD correction $\delta$ in the
MSSM as a function of $M_S$, taking $A_t=800~GeV$, $M_2=110~GeV$,
$m_{A^0}=270~GeV$, $\mu=-200~GeV$ and $\tan\beta=6$. The figure
demonstrates that the relative NLO QCD corrections in the MSSM at
the Tevatron and the LHC, are stable when $M_S$ changes from
$400~GeV$ to $2~TeV$. Their values are about $25\%$ at the LHC and
$-18\%$ at the Tevatron. We find from our calculation that when
$M_S$ is taken as a large value, the correction from the NLO SUSY
QCD correction part decreases to a constant due to the decouple of
heavy stop quarks.
\par
Fig.7 shows the total QCD relative correction $\delta$ in the MSSM
as a function of $m_{\tilde{g}}$ with the input parameters same as
in Fig.4. We can see from this figure that the $\delta$ have a
concave structure in the vicinity of $m_{\tilde{g}} \sim 140 -
150~GeV$, where the masses satisfy the relation $m_{\tilde{g}}+m_t
\approx m_{\tilde{t}_1}+m_{h^0}$ and the re-scattering enhancement
$\tilde{t}_1^* \to \tilde{g}+t \to \tilde{t}_1+h^0$ takes place.
When $m_{\tilde{g}}$ goes from $400~GeV$ to $2000~GeV$, the
relative corrections are very stable, they are about $24\%$ for
the LHC and $-18\%$ for the Tevatron. Similar with the case in
Fig.6, due to the decouple effect the correction of the SUSY QCD
correction part decreases to a constant when $\tilde{g}$ is
getting heavy.
\par
Fig.8 presents the total NLO QCD relative correction $\delta$ in
the MSSM as a function of the SUSY parameter $A_t$, assuming
$M_S=400~GeV$, $M_2=110~GeV$, $m_{A^0}=270~GeV$, $\mu=-200~GeV$
and $\tan\beta=6$. We can see from the figure that the total NLO
QCD relative corrections in the MSSM are very sensitive to $A_t$
in the region near the position of $A_t=1000~GeV$(where
$m_{\tilde{t}_1}=108.6~GeV$). Actually, the reason for that is
because of the large contribution from the light stop quark
$\tilde{t}_1$ loops. When the chosen parameters $A_t$ and $\mu$
make a large mass splitting between the two scalar top-quarks,
then the $\tilde{t}_1$ becomes light. We can read from the figure
the total NLO QCD relative correction $\delta$ in the MSSM can
reach $-24\%$ at the Tevatron and $7\%$ at the LHC when $A_t$ is
near $1000~GeV$.
\par
In Fig.9, we show the total NLO QCD relative correction $\delta$
in the MSSM as a function of the SUSY parameter $\mu$, assuming
$A_t=800~GeV$, $M_S=400~GeV$, $M_2=110~GeV$, $m_{A^0}=270~GeV$ and
$\tan\beta=6$. We can see that the total NLO QCD relative
corrections in the MSSM increase slowly when $\mu$ goes up from
$-1000~GeV$ to $1000~GeV$, this is because the absolute values of
the negative corrections from the SUSY QCD part are becoming
smaller as $\mu$ increases. The value of $\delta$ at the Tevatron
can be beyond $-22\%$ when $\mu$ is about $-1000~GeV$.
\par
In this paper we calculated the NLO QCD corrections to the
processes $p\bar{p}/pp \to t\bar{t} h^0+X$ in the MSSM at the
Tevatron and the LHC. We analyzed the dependence of the corrected
cross sections or relative corrections on the
renormalization/factorization scale $Q$, SUSY parameters
$\tan\beta$, $M_S$, $m_{\tilde{g}}$, $A_t$ and $\mu$,
respectively. It shows that the dependence of the total NLO QCD
corrected cross section in the MSSM on the scale $Q$ is
significantly reduced comparing with the $\sigma_{LO}$. With our
chosen parameters, the numerical results demonstrate that the
relative correction is obviously related to $m_{\tilde{g}}$, $A_t$
and $\mu$ in some parameter regions, but not very sensitive to
$\tan\beta$, $M_S$ at both the Tevatron and the LHC for our
specified parameters. We conclude that the total NLO QCD
corrections are generally moderate, which have the values in the
range of few percent to about $20\%$ in most of the SUSY parameter
space. We find that the relative correction from the NLO SUSY QCD
correction part becomes to be constant when either $M_S$ or
$m_{\tilde{g}}$ has large value. We find also the relative
correction of the SUSY QCD part will be largely enhanced when the
mass splitting between stop-quarks is large, the total NLO QCD
relative correction in the MSSM $\delta$ can reach $-24\%$ at the
Tevatron and $7\%$ at the LHC.

\paragraph{Acknowledgments}
This work was supported in part by the National Natural Science
Foundation of China and special fund sponsored by China Academy of
Science.

\begin{figure}[htp]
\centering
\includegraphics[height=3.5in]{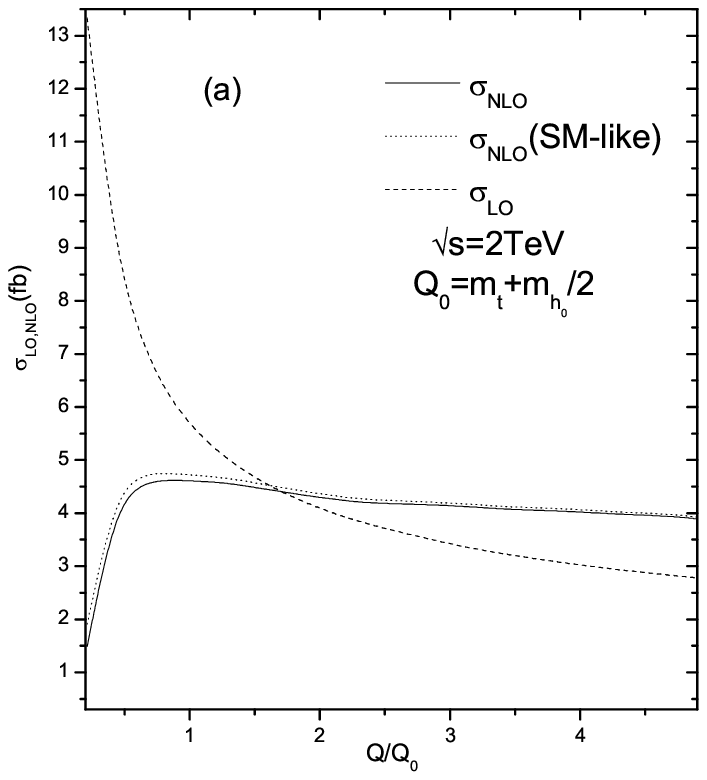}%
\hspace{0in}%
\includegraphics[height=3.5in]{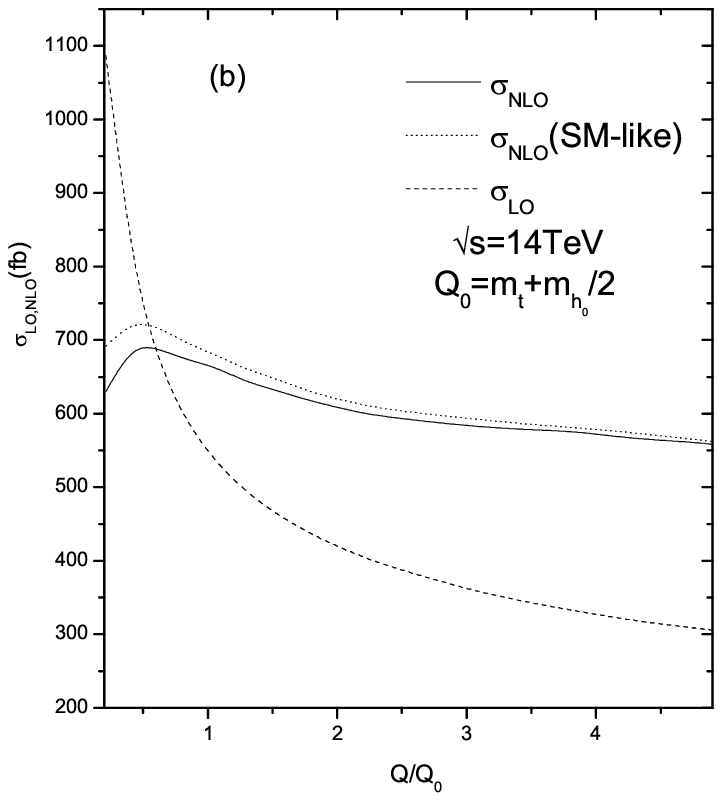}
\caption{The cross sections $\sigma_{LO}$ at the leading order and
$\sigma_{NLO}$ involving the NLO QCD corrections in the MSSM as
the functions of the renormalization/factorization scale $Q$ with
$m_{\tilde{g}}=300~GeV$ and the other parameters are from FormCalc
by using the input SUSY parameters: $M_S=400~GeV$, $M_2=110~GeV$,
$A_t=800~GeV$, $m_{A^0}=270~GeV$, $\mu=-200~GeV$ and
$\tan\beta=6$. Fig.4(a) is for the process $p\bar p \to t\bar
th^0+X$ at the Tevatron and Fig.4(b) for the process $pp \to t\bar
th^0+X$ at the LHC.}
\end{figure}

\setlength{\abovecaptionskip}{0pt}
\begin{figure}[htp]
\centering
\scalebox{1}{\includegraphics[170,275][180,280]{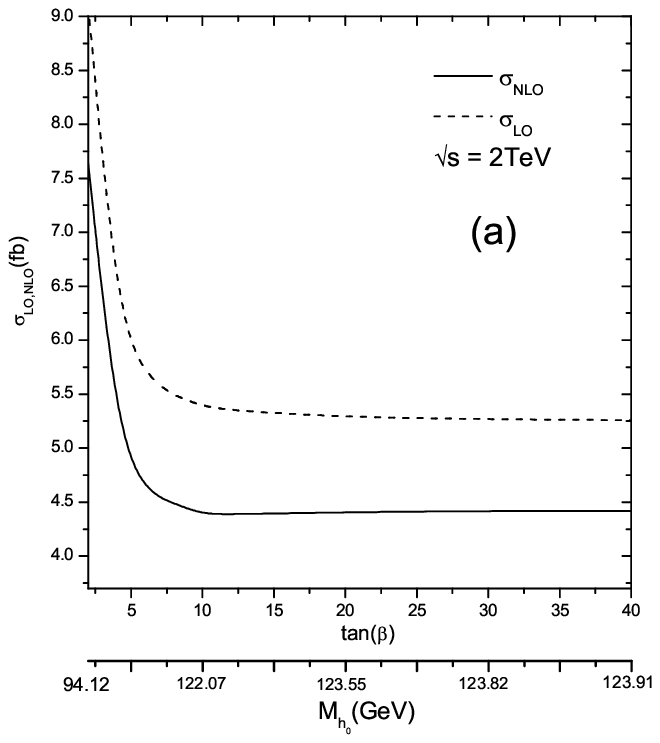}}
\scalebox{1}{\includegraphics[-20,52][120,280]{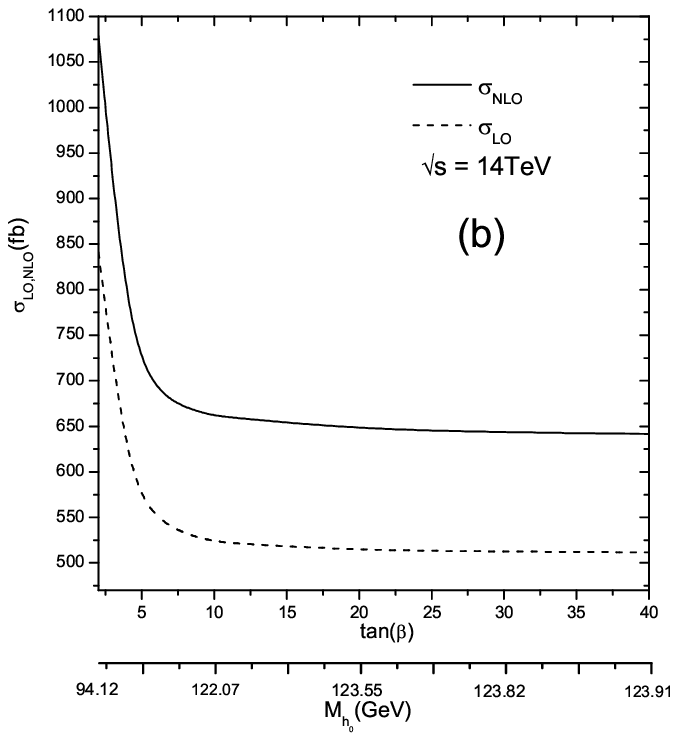}}
\scalebox{0.9}[0.8]{\includegraphics*[-110,23][450,280]{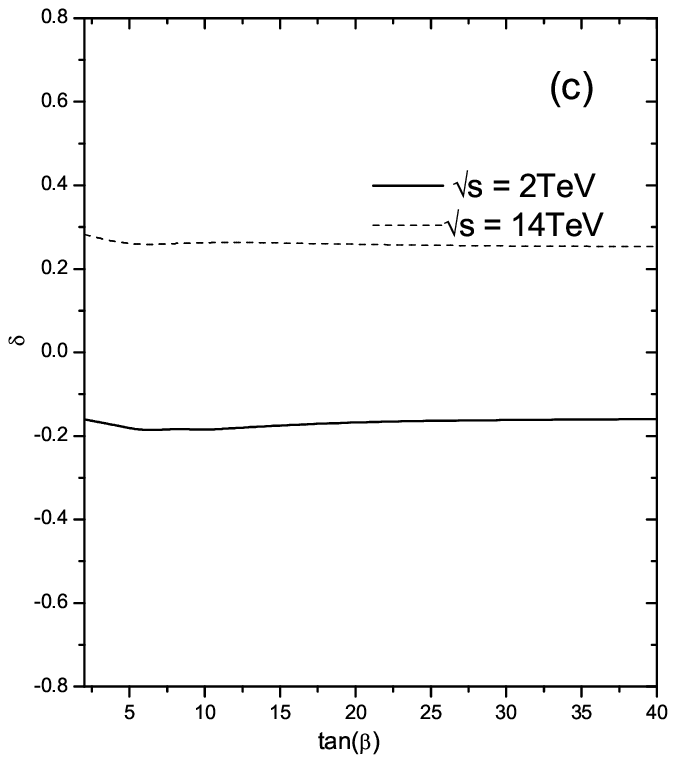}}
\center{\caption{ The cross sections $\sigma_{LO}$ at the leading
order and $\sigma_{NLO}$ involving the NLO QCD corrections in the
MSSM as the functions of $\tan\beta$ with the input parameters
$A_t=800~GeV$, $M_2=110~GeV$, $m_{A^0}=270~GeV$, $M_S=400~GeV$,
$\mu=-200~GeV$. Fig.5(a) is for the process $p\bar p \to t\bar
th^0+X$ at the Tevatron and Fig.5(b) for the process $pp \to t\bar
th^0+X$ at the LHC. Fig.5(c) shows the relative NLO QCD correction
as the function of $\tan\beta$ in both the Tevatron and LHC.}}
\end{figure}

\setcaptionwidth{2.7in}
\setlength{\abovecaptionskip}{-10pt}

\begin{figure}[htp]
\centering
\begin{minipage}[c]{0.5\textwidth}
\includegraphics[width=3.2in,height=3in]{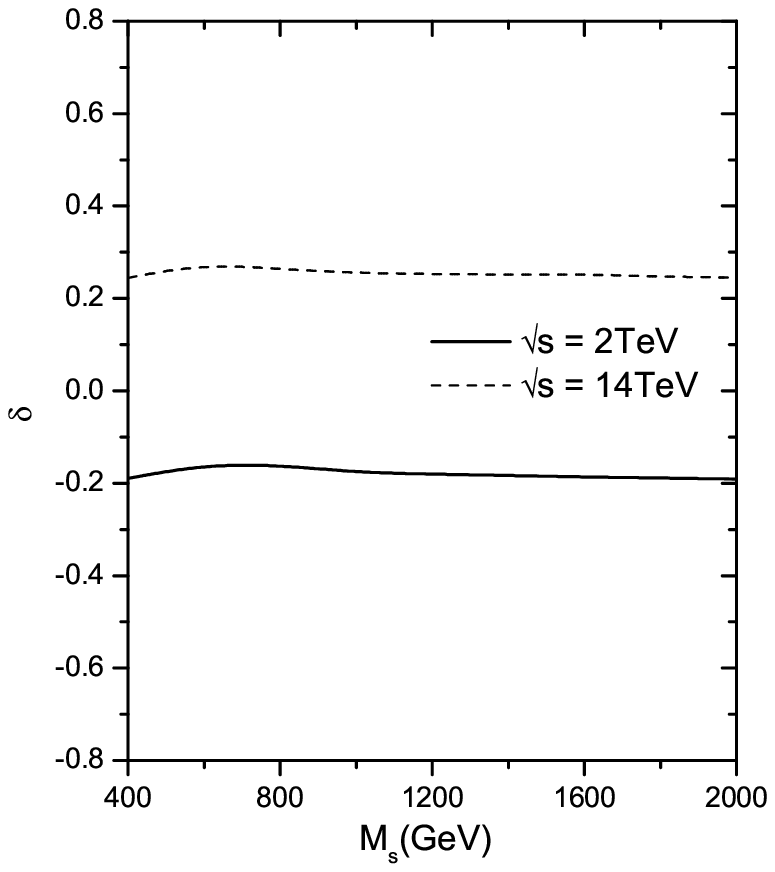}
\caption{The total NLO QCD relative corrections($\delta$) of the
processes $p \bar{p}/pp \to t \bar{t} h^0+X$ at the Tevatron and
the LHC, as the functions of $M_S$.}
\end{minipage}%
\begin{minipage}[c]{0.5\textwidth}
\includegraphics[width=3.2in,height=3in]{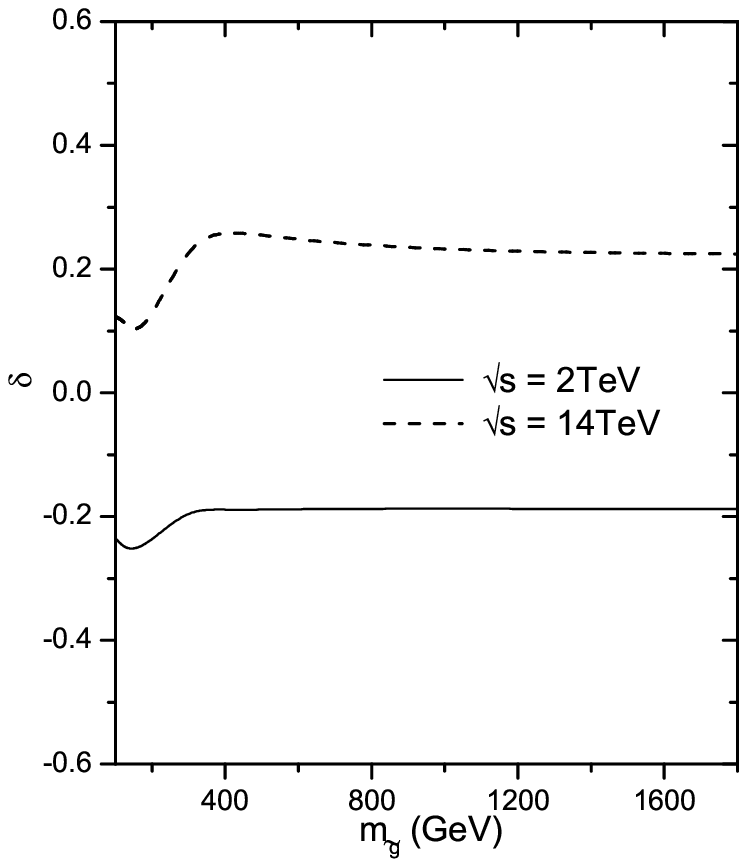}
\caption{The total NLO QCD relative corrections($\delta$) of the
processes $p \bar{p}/pp \to t \bar{t} h^0+X$ at the Tevatron and
the LHC, as the functions of $m_{\tilde{g}}$.}
\end{minipage}
\begin{minipage}[c]{0.5\textwidth}
\includegraphics[width=3.2in,height=3in]{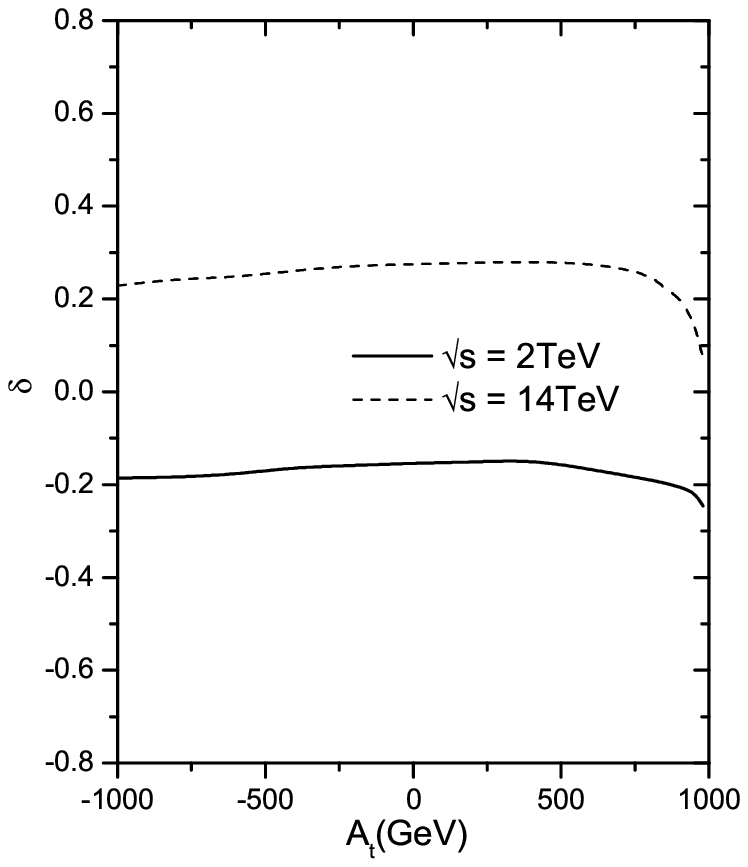}
\caption{The total NLO QCD relative corrections($\delta$) of the
processes $p \bar{p}/pp \to t \bar{t} h^0+X$ at the Tevatron and
the LHC, as the functions of $A_t$.}
\end{minipage}%
\begin{minipage}[c]{0.5\textwidth}
\includegraphics[width=3.2in,height=3in]{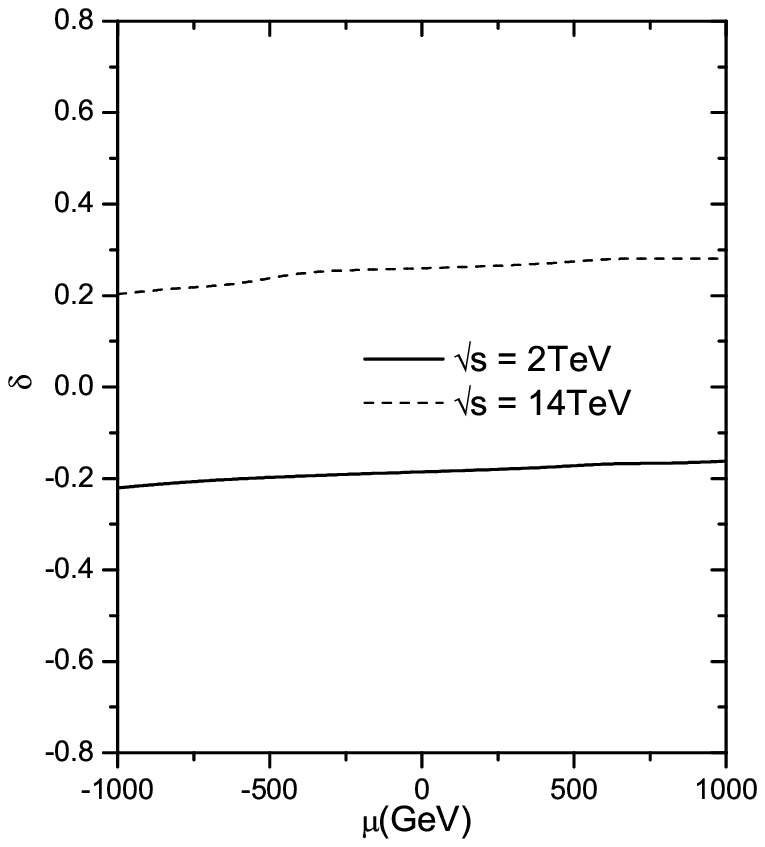}
\caption{The total NLO QCD relative corrections($\delta$) of the
processes $p \bar{p}/pp \to t \bar{t} h^0+X$ at the Tevatron and
the LHC, as the functions of $\mu$.}
\end{minipage}%
\end{figure}

\end{document}